\documentclass[aps,pra,twoside,twocolumn,10pt]{revtex4-2}
\usepackage[colorlinks=true, citecolor=blue, urlcolor=blue ]{hyperref}
\usepackage{epsfig,newlfont,amssymb,amsfonts,amsmath,bm,subfigure,palatino,mathtools,amsthm,braket,soul,enumitem,color,graphics,graphicx,times,physics}
\usepackage[normalem]{ulem}
\usepackage{multirow}
\usepackage{array}

\newcolumntype{P}[1]{>{\centering\arraybackslash}p{#1}}
\graphicspath{{Figures/}{Dynamics/}}

\begin{document}
\title{
Quantum-enhanced sensing  with variable-range interactions 
}
\author{Monika Mothsara$^{1,2,3}$, Leela Ganesh Chandra Lakkaraju$^{2,4,5}$, Srijon Ghosh$^2$,   Aditi Sen(De)$^2$}
\affiliation{$^1$ Department of Physical Sciences, IISER Berhampur, Berhampur - 760010, India}
\affiliation{$^2$ Harish-Chandra Research Institute, A CI of HBNI, Chhatnag Road, Jhunsi, Prayagraj - 211019, India}
\affiliation{$^3$ Atominstitut, Technische Universität Wien, Vienna Center for Quantum Science and Technology (VCQ), Stadionallee 2, 1020 Vienna, Austria}
\affiliation{$^4$Pitaevskii BEC Center and Department of Physics, University of Trento, Via Sommarive 14, I-38123 Trento, Italy} 
\affiliation{$^5$ INFN-TIFPA, Trento Institute for Fundamental Physics and Applications, Trento, Italy}

\begin{abstract}

The typical bound on parameter estimation, known as the standard quantum limit (SQL), can be surpassed by exploiting quantum resources such as entanglement. To estimate the magnetic probe field, we propose a quantum sensor based on a variable-range many-body quantum spin chain with a moderate transverse magnetic field. We report the threefold benefits of employing a long-range system as a quantum sensor. First,  sensors with quasi-long-range interactions can always beat the SQL for all values of the coordination number, while a sensor with long-range interactions does not have this ubiquitous quantum advantage. Second, a long-range Hamiltonian outperforms a nearest-neighbor (NN) Hamiltonian in terms of both estimating precision and system-size scaling. Finally, we observe that the system with long-range interactions can go below the SQL in the presence of a high temperature of the initial state, while sensors having 
NN interactions cannot.  Furthermore, a sensor based on the long-range Ising Hamiltonian proves to be robust against impurities in the magnetic field and when the time-inhomogeneous dephasing noise acts during interaction of the probe with the system.

\end{abstract}

\maketitle

\section{Introduction}
\label{sec:intro}

Gathering information about any physical quantity or parameters involves repeated measurements on the system, which inevitably introduces errors in estimation. In a classical domain, the central limit theorem ensures that for repeated measurements on the parameters, the error is proportional to  $\nu^{-1/2}$, known as the standard quantum limit (SQL) \cite{Sensing_RMP_2017}, where $\nu$ denotes the total number of measurements performed in the system.  However, in a quantum world,  it has been realized that the existence of entanglement \cite{HHHH_entanglement} in the composite system is capable of altering the performance of quantum information processing tasks, including quantum communication and computation \cite{ekert1991, bennett1992, bennett1993, Rausendorf2001}. In a similar spirit, it was discovered that by leveraging quantum resources like entanglement or squeezed states \cite{wineland1992, kitagawa1993}, the precision in parameter estimation might further be increased, lowering the error's scaling to the so-called Heisenberg limit (HL) of $\nu^{-k}$  for the sensing of $k$ parameters \cite{multiparameter_caves_2007, giovenneti2006}.

To estimate the strength of the magnetic field interacting with a quantum system, quantum spin systems consisting of $N$ subsystems as probes are routinely employed \cite{metro_manybody_critic_zanardi_2008, metro_manybody_critic_paris_2008, manybody_metro_critic_compressed_dur_2016, manybody_metro_critic_paris_2016, manybody_metro_critic_horodecki_2018, manybody_metro_critic_PRR_2023, manybody_metro_critic_PRL_2023}, which are then exposed to an external magnetic field for a certain period. 
As a result of the finite energy shifts caused by the magnetic field, the system acquires a relative phase, allowing the corresponding parameters to be estimated by comparing the input and output of the probe after performing measurements \cite{giovenneti2006, giovannetti_nature}. 
Due to the Cram{\'e}r-Rao bound \cite{cramer1946mathematical}, the standard deviation in the estimation of the magnetic field for $N$-body quantum sensors is lower bounded by the inverse of the square root of the Fisher information. 
Entangled states, in particular, path entangled $NOON$ states \cite{kok2002} and   Greenberger-Horne-Zeilinger (GHZ) states \cite{ghz}, help accomplish more precise parameter estimation, outperforming the existing precision limitations, $N^{-1/2}$ of utilizing uncorrelated probes \cite{caves1981}. In addition, quantum control  \cite{ landau_zenar_metrology_2017, quantum_control_campo_2022} and sequential measurements \cite{abolfazal_sequential_2022} allow better scaling in time to achieve minimum uncertainty. 

Thus, a system governed by a many-body interacting spin Hamiltonian which typically possesses a significant amount of non-local correlations, e.g., entanglement, can serve as an efficient probe in quantum metrology \cite{metro_manybody_critic_zanardi_2008}. 
It has also been demonstrated that when the system is fully accessible, preparing the probe's initial state at the critical point, where the correlation length diverges, can be exploited to increase the precision limit \cite{zanardi2006, metro_manybody_critic_zanardi_2008, zanardi2007,  metro_manybody_critic_paris_2008, manybody_metro_critic_horodecki_2018}. However, such a benefit disappears when the subsystems of the entire system are used which can be remedied by evolving the system periodically \cite{utkarsh2021}. 
Moreover, recent work shows that the vulnerability of a near-transition of a quantum many-body system from localization to delocalization provides a resource for achieving quantum-enhanced sensitivity in the parameter estimation protocol \cite{sahoo2023}. Towards minimizing the environmental impact on the system, it was suggested \cite{Matsuzaki_2021} that a chain of qubits with only nearest neighbor interaction may be used to construct a quantum sensor in which measurements are made rather than interactions between the sites being modulated. 
Importantly, quantum advantages obtained in the parameter estimation have been demonstrated in the laboratories using trapped ions \cite{trapped_ion_sensing1, trapped_ion_sensing2}, thermal vapors of atom \cite{vapour_sensing1, thermal_vapour_sensing}, nuclear magnetic resonance \cite{nmr_sensing}, cold atoms \cite{cold_atom_sensing}, nitrogen-vacancy sold-state qubit \cite{nv_sensing1} and superconducting qubits \cite{superconducting_qubit_sensing}.

It is always fascinating to comprehend how the performance of a quantum sensor is influenced in the presence of noise, since setting up and manipulating the isolated system is the ideal situation.
Although in the presence of Markovian dephasing noise, quantum sensors can not beat the SQL \cite{huelga, open_metro_review_2016}, the non-Markovian dynamics of entangled initial states are shown to be better than the unentangled ones, thereby providing quantum advantage \cite{jones2009, matsuzaki2011, chin2012, katar2015, bhattacharyya2022, bhattacharyya2023}. To circumvent the detrimental effect of noisy environments on quantum sensors, numerous mechanisms such as dynamical decoupling \cite{llyod1999, Pham2012, Yang_2022}, error correcting codes \cite{lukin2014}, and error-mitigation by purification \cite{error_purfication_prl_2022} are being developed to reach the Heisenberg limit. 

All the aforementioned works have been carried out using uncorrelated or correlated systems, considering only the nearest-neighbor (NN) interactions between subsystems (see Ref. \cite{abolfazl_longrange_2023} for a recent study on the scaling of quantum Fisher information in long-range interacting systems). However, beyond NN interactions, i.e., long-range interactions naturally arise in physical systems like trapped ions \cite{PhysRevLett.98.253005,gerritsma2010quantum,ion_trapp_review_2021} and ultracold atoms in optical lattices \cite{cramer2013spatial,mandel2003controlled, non_equi_lr1,non_equi_lr2}. The long-range interaction corresponds to a dipolar interaction between spins, governed by Coulomb forces. For example, in nuclear magnetic resonance systems to enhance entanglement-based metrology \cite{nmr_metrology_2005},  a single spin, referred to as the target spin (typically protons of the carbon nucleus), interacts with a group of spins known as amplifier spins, which are connected via long-range dipolar interactions among the amplifier spins. It was shown to enable precise estimation of the magnetization of the target spin. Given such a precedent in the application of long-range systems in metrology and the availability of several other platforms in simulating out-of-equilibrium long-range systems, we propose a quantum sensor based on a quantum many-body system involving variable-range interactions, which can be simulated in a currently available technology. Specifically, the canonical equilibrium state of the $XY$ model with a variable-range interaction in the presence of a magnetic field is employed as the initial state of the quantum sensor. Instead of tuning the interaction between the sites \cite{matsuzaki2011}, an optimal single-qubit measurement is performed at one of the endpoints of the spin chain, inducing a flipping mechanism in the entire spin chain during the dynamics of the system, which results in a GHZ state with high fidelity. 

The variable-range interacting systems are studied in various contexts. Specifically, in a seminal work by Majumdar and Ghosh \cite{Majumdar1969Aug}, it was discovered that by introducing next-nearest-neighbour interactions in the Heisenberg model, known as the $J_1-J_2$ model, one can obtain a gapped phase which was not present in the parent model. Importantly, such an interacting Hamiltonian can be found in solid state systems such as $CuGeO_3$ \cite{solid_state_1_nnn} and $NaV_2O_5$ \cite{solid_state_2_nnn} and can also be engineered in a quantum simulator \cite{Zeilinger_four_qubit_next_nearest, J1_J2_simulator}, photonic crystal waveguide array \cite{photonic_nnn_arxiv} and many more systems. In the same spirit, the next-next-nearest neighbor interaction is added to a $J_1-J_2$ model in order to explain the phenomena of magnetic features of layered compounds in the transition metals \cite{J3_original_1, J3_original_2, J3_original_3}. These interactions occur naturally in solid-state compounds such as $Na_2 Ir O_3$ and $Li_2 Ir O_3$ \cite{J1_J2_J3_solid_state}. All these models with variable-range interactions can be realized in a quantum simulator due to the extraordinary progress of the cold-atom setup, which ensures the realization and manipulation of many-body system in a controlled manner \cite{mandel2003controlled}. Given such extensive previous implementations, one can expect that the advantage observed using the variable-range Hamiltonian can be experimentally realized in currently available technologies.

To precisely estimate the strength of the probe's magnetic field, the system strongly interacts with the probe for a sufficient amount of time before performing another optimal measurement on a single-qubit for read-out purposes after turning off the magnetic field to be estimated and evolving the system for a suitable time period. It is important to note that the performance of the sensing protocol is critically dependent on several interrelated system parameters such as the time of initial evolution, the measurement setting to create the GHZ state and at the final step for read out, the time in which the probe state interacts with the system, the strength of the magnetic field of the quantum sensor's initial state, and the range of interactions. To begin, we observe that the performance of the quantum sensing protocol used here is most effective when the initial Hamiltonian is the transverse Ising model. We uncover the measurement operators that lead to high fidelity and sensitivity by optimizing over single-qubit rank-$1$ measurement operators.  Furthermore, the time span during which the system should evolve is specified.

The entire examination can provide the platform for addressing the impact of a variable range of interactions on the sensitivity. First of all, the range of interactions and coordination number of the initial Hamiltonian are closely connected. We specifically report that when the initial state is created involving the quasi-local interactions, quantum sensors can exhibit improved sensitivity by surpassing the SQL, and this is true for any coordination number. More significantly, the acquired sensitivity is more than that which can be attained using nearest-neighbor interactions, and this benefit of the quasi-long-range Hamiltonian persists even in the presence of a high temperature.  However, in the case of sensors having long-range interactions, there exists a critical coordination number over which the sensitivity falls below the SQL, exhibiting that the range of interactions can also be deleterious. We also determine that when the magnetic field strength is moderate, the long-range model can outperform the NN model. Further, we find that the sensor with long-range interaction can provide a benefit over the SQL with the increase of system size. We demonstrate that the benefit in sensitivity obtained for the quasi-local  Hamiltonian persists even in the presence of disorder in the transverse magnetic field and non-Markovian-type dephasing noise.

The paper is organized as follows: Sec. \ref{sec:strategy} deals with the quantum sensing protocol adopted here using a variable-range Hamiltonian. In Sec. \ref{advantage}, we analyze the performance of a long-range interacting system as a potential candidate for quantum sensors. The beneficial effect of a higher coordination number instead of only the nearest neighbor scenario is also discussed, indicating the effectiveness of using such systems as quantum sensors. The robustness in the estimated uncertainty due to the inevitable presence of impurities in the system is addressed in Sec. \ref{sec:disorder} while we investigate the effect of local dephasing noise in the estimation protocol in Sec. \ref{decoherence}. Finally, in Sec. \ref{conclusion}, we summarize the results. 

\section{Strategy of estimating parameter in quantum spin models}
\label{sec:strategy}

We present a parameter estimating protocol involving a quantum spin chain, motivated by the recently introduced scheme \cite{Matsuzaki_2021}, which is markedly different from the conventional one \cite{manybody_metro_critic_paris_2016, metro_manybody_critic_zanardi_2008}. Using a variable-range quantum spin chain as the quantum sensor, we systematically demonstrate its enhanced performance in estimating the external probe, i.e., the magnetic field $H_{probe}$, without controlling the interaction between the qubits and also by performing a local measurement at the initial stage of the protocol and at the time of readout \cite{Matsuzaki_2021}.  

\subsection{Parameter estimation scheme.} Precisely, we consider a variable-range interacting Hamiltonian, $H$, consisting of $N$ spin-$1/2$ particles, as the quantum sensor for the parameter estimation protocol. The total sensor Hamiltonian consists of two parts
\begin{equation}
    H = H_{field} + H_{int},
    \label{total_H}
\end{equation} where $H_{field}$ and $H_{int}$ represent local magnetic field and interaction terms respectively. The exact forms considered in this work will be presented later in this section. 

\textit{Step 1. Preparation of a suitable state for sensing.} The initial state of the sensor is prepared as the canonical equilibrium state of $H$ at inverse temperature $\beta = \frac{1}{k_B \tau}$, with $k_B$ the Boltzmann constant, and \(\tau\) the temperature, i.e., $\rho_{th}(0) = \frac{\exp(-\beta H)}{Z}$, where $Z = \mbox{Tr}[\exp(-\beta H)]$ represents the partition function of the system. A single qubit arbitrary projective measurement is performed on the first spin of the chain to initiate the \emph{flipping} mechanism in the entire system. They are written as $ P_{1}(\theta, \phi) = |\mu \rangle \langle \mu|$ and $ P_{1}^\perp(\theta, \phi) = |\mu^{\perp} \rangle \langle \mu^{\perp}|$ with $|\mu\rangle = \left(\cos\frac{\theta}{2} |0\rangle + e^{i\phi} \sin\frac{\theta}{2} |1\rangle \right)$, $|\mu^\perp \rangle  = \left(\sin\frac{\theta}{2} |0\rangle - e^{i\phi} \cos\frac{\theta}{2} |1\rangle \right)$  such that $0 \le \theta \le \pi$ and $0 \le \phi \le 2\pi$. By optimizing over $\theta$ and $\phi$ in the measurement, the state obtained after measurement evolves with a global unitary operator, \(U_{t*}\), governed by the initial Hamiltonian $H$ for a time interval $t^{*}$, given by $e^{-iHt^*}$. Since it is known that $(N-1)$ party GHZ state, $|GHZ\rangle = \frac{1}{\sqrt{2}}(|0\rangle^{\otimes N-1} + |1\rangle^{\otimes N-1})$ possesses all the characteristics of being a resource for quantum metrology \cite{wineland1992,PhysRevLett.65.1838}, we maximize the fidelity between the output state obtained after evolution and the $(N-1)$ party GHZ state, over the parameters involved in measurements, $\theta$ and $\phi$. Mathematically, we compute 
\begin{align}
    \mathcal{F} = \max_{\theta, \phi} \langle GHZ &| \text{Tr}_N( e^{-iHt^*} (P_1 (\theta, \phi) \otimes I_{N-1}) \rho_{th}(0) \nonumber \\ 
    &(P_1(\theta, \phi) \otimes I_{N-1})  e^{iHt^*}  )|GHZ \rangle
    \label{eq:fidelity}
\end{align} 
 where $I_{N-1}$ denotes the identity operator in $2^{N-1}$ dimension and \(\text{Tr}_N\) denotes the tracing out of a single site at the edge on which the measurement is not performed, which e.g. can be the site, \(N\). 

\textit{Step 2. Probe in action.} Let us now allow the probe magnetic field to interact with the sensor, by turning off the initial external magnetic field, thereby leading to the evolution of the system under the unitary operator generated by $H_{probe}$ and $H_{int}$ for a time interval $t_{int}$, i.e., $ U_{t_{int}} = e^{-i(H_{probe}+H_{int})t_{int}}$.  Hence, the dynamical state at $t=t_{int}$ contains the information about the probe. Again, one switches off the probe Hamiltonian and evolves the system according to $ e^{-iHt^*}$ for a fixed time, $t^*$. We apply the additional unitary $U_{t^*}$ to disentangle the state and convert it back to a state in which information about the target field is concentrated, which is useful for the readout. Hence, the unitary operators that act on the system for the entire protocol read as
\begin{equation}
    U_{total} = U_{t^{*}} U_{t_{int}} U_{t^{*}}.
\end{equation}

\textit{Step 3. Read out.} Finally, the information about the probe field can be achieved with high precision by measuring the first qubit using an arbitrary projector, $ \{P_1 (\theta^\prime, \phi^\prime), P_1^\perp (\theta^\prime, \phi^\prime)\}$. It is important to note that instead of performing the measurement on each qubit of the entire system or the global measurement on the whole, we only require measurement of the first qubit of the quantum sensor since all the necessary phase information is concentrated in that qubit via $U_{t_{int}}$. Thus this approach simplifies the measurement process and the cost of the measurement. Again, the optimization is performed over the parameters to maximize the estimation of the probe (see Appendix \ref{app:secular} for an example that clearly illustrates the benefit of including $U_{t^*}$ in this scheme).

\label{method_protocol}

\subsection{A variable-range spin model as a sensor}

\begin{figure}
    \centering
    \includegraphics[width=\linewidth]{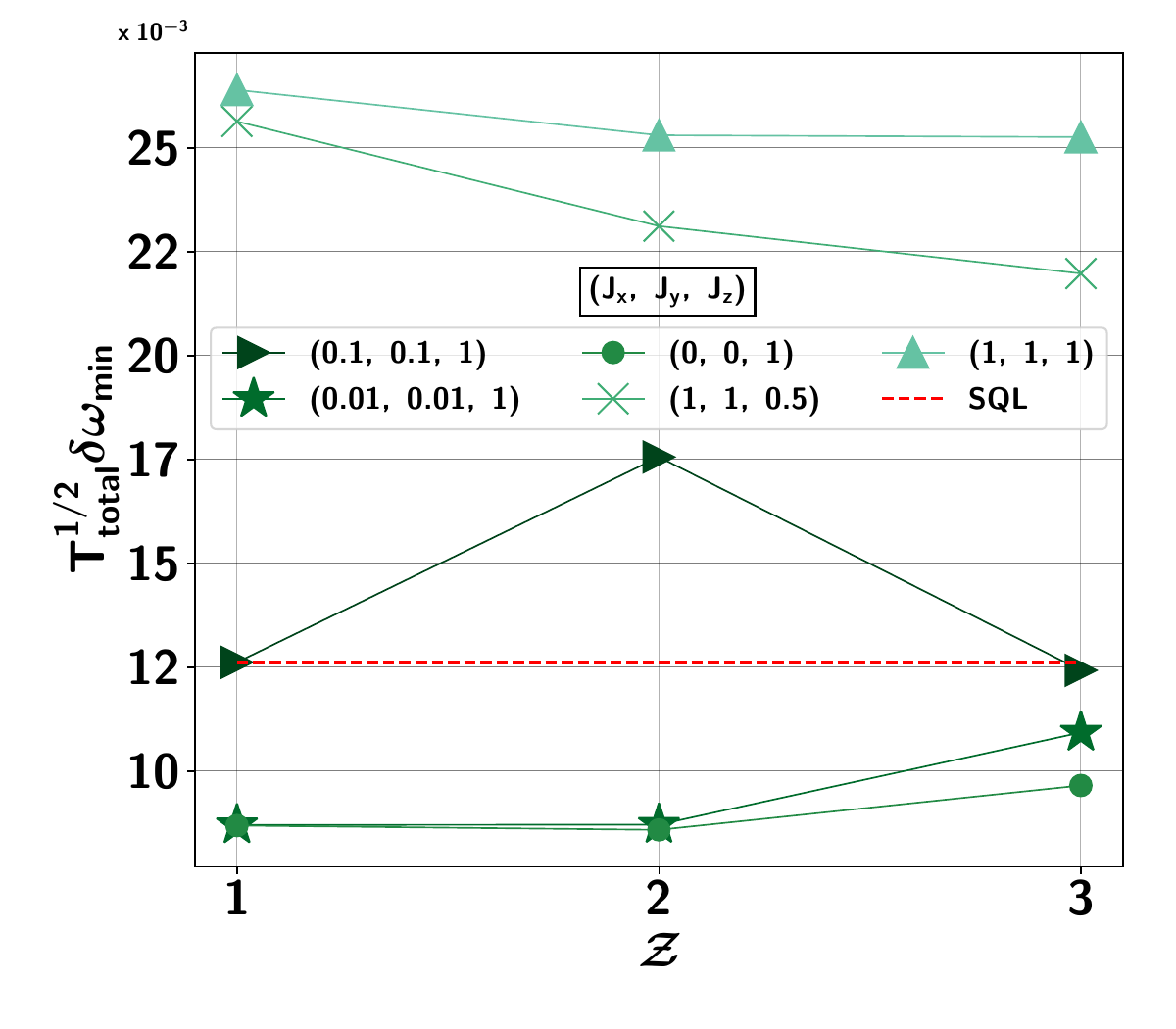}
    \caption{ (Color online) Minimum uncertainty, \(\sqrt{T_{total}}\delta \omega_{\min}\) (ordinate) vs coordination number $\mathcal{Z}$ (abscissa). Different lines correspond to different sets of \((J^x, J^y, J^z)\). When the values of $J^x$ and $J^y$ are decreasing, along with keeping $J^z$ constant, the value of uncertainty falls below the SQL. This possibly indicates that the long-range Ising model (Eq. \ref{H_tfi}) is better than the $XYZ$ model as a sensor. Both axes are dimensionless. }
    \label{fig:j_x}
\end{figure}

To illustrate the advantages of constructing a quantum sensor with a long-range quantum spin chain, we consider an $XYZ$ chain with variable-range interactions in the presence of a transverse magnetic field as our workhorse, represented as
\begin{align}
    H_{xyz}&=-\sum_{\substack{i<j \\|i-j|\leq\mathcal{Z}}}^{N-1} \frac{J_{i j}}{4}\left(J^x \sigma_i^x \sigma_{j}^x+J^y \sigma_i^y \sigma_{j}^y + J^z \sigma_i^z \sigma_{j}^z \right)   \nonumber \\
    &+ \frac{h^x}{2} \sum_{i=1}^N  \sigma_i^x
    \label{eq:xy_model}
\end{align}
where $N$ is the total number of qubits and $\sigma^{k}$ $(k = x, y, z)$ are the usual Pauli matrices. Here $h^{x}$ is the strength of the magnetic field applied in the $x$-direction, representing $H_{field}$.   The first term of $H_{xyz}$ is the interaction Hamiltonian, $H_{int}$, in which $J_{ij}$ denotes the strength of the interaction between any two sites, $i$ and $j$ which depend on the distance between the sites $i$ and $j$ and the corresponding power law decay, $J_{ij}\sim A^{-1}|i-j|^{-\alpha}$, with $\alpha$ the fall-off rate and $A = \sum_{|i-j|=1}^{N-1} |i-j|^{-\alpha}$ being the normalization constant, known as Kac factor \cite{kac1963van}. Here $J_{ij} > 0$ represents the strength of the ferromagnetic interactions between spins and $\mathcal{Z}$ is the coordination number (range of interaction). In addition, $J^x, ~ J^y,  \text{ and } J^z \in [0,1]$ denote the strengths of the interaction in the $xyz$ plane. By optimizing them over the said range, we find out that in order to obtain the minimum uncertainty in the parameter estimation scenario, the optimal Hamiltonian in this class is the Ising Hamiltonian, i.e., $J^x = J^y = 0 \text { and } J^z \ne 0$ (see Fig. \ref{fig:j_x}).  Therefore, substituting $J^x = J^y = 0 \text { and } J^z \ne 0$ in Eq. (\ref{eq:xy_model}), the sensor Hamiltonian, representing the transverse Ising long-range model, reads 
\begin{equation}
H = -\sum_{\substack{i<j \\|i-j|\leq\mathcal{Z}}}^{N-1} \frac{J_{i j}}{4} \sigma_i^z\sigma_j^z + \frac{h^x}{2} \sum_{i=1}^N  \sigma_i^x.
\label{H_tfi}
\end{equation}
Furthermore, the probe field is taken to be
\begin{equation} 
H_{probe}= \frac{\omega}{2} \sum_{i=1}^N  \sigma_i^z,
\label{H_w}
\end{equation}
where $\omega$ is the strength of the magnetic field to be estimated. Additionally, the total magnetization density, in the $z$-direction, is defined as 
\begin{equation}
    \frac{M^{z}}{N} \equiv \frac{\langle \sigma^{z}\rangle}{N} := \frac{1}{2N}\sum_{i = 1}^{N} \sigma^{z}_{i},
    \label{magnetization}
\end{equation} 
which is computed to obtain the optimal $t^*$ required for the minimum value of uncertainty (defined below) of the estimated parameter $\omega$.

Let us define how to precisely estimate the frequency parameter $\omega$ \cite{Sensing_RMP_2017,PhysRevA.99.022325,huelga}. 
For a single qubit, the uncertainty (error) in the estimation can be expressed as \footnote{Note that in our analysis, we consider both $p$ and $1-p$ for computing $\delta \omega$, ensuring that the measurement process remains fully inclusive.}
\begin{equation} \label{uncertainty}
\delta \omega = \frac{\sqrt{p(1-p)t_{int}}} {\left|\frac{\partial p}{\partial \omega}\right|\sqrt{T_{total}}},
\end{equation}
where $p = \mbox{Tr}(P_1(\theta^\prime, \phi^\prime)U_{total}P_1(\theta, \phi)\rho_{th}(0) P_1(\theta, \phi)U_{total}^\dagger)$.
Here, $T_{total}$ and $t_{int}$ represent the overall duration of the experiment and the time during which the system evolves under the target magnetic field, respectively.  
By estimating the frequency parameter, we can calculate the amplitude of the target magnetic field. Here, we are assuming $t_{int}$ to be significantly longer than other time scales of the procedure. The main aim of quantum sensing is to minimize the errors in the protocol, thereby maximizing the estimation of parameters, \(\omega\). The performance quantifier for the protocol is given by $$\delta \omega_{\min} = \min_{t^* < t^*_{search}} \delta \omega,$$ where $t^*_{search}$ is the maximum time until which the minimum uncertainty is searched. We describe the dependence of system parameters on the value of $t^*_{search}$ in the following section. Due to the nonlinear nature of the landscape of the parameters involved in the system under study,  we opt for a nonlinear optimization library known as NLopt \cite{NLopt}.

\section{Beneficial role of variable-range interaction in quantum metrology}
\label{advantage}

Based on the sensing protocol prescribed in the preceding section, we now analyze the performance of a quantum sensor constructed using the long-range transverse Ising model to estimate the uncertainty in determining the strength of the probe magnetic field. For systematic discussion, let us first define some relevant time scales during the protocol apart from the aforementioned evolution time. In the course of finding the minimum uncertainty of the estimated parameter, $t^*_{opt}$ is that particular time for which the evolved state is closest to the $GHZ$ state, thereby providing the maximum advantage in the estimation scenario. To determine this particular time instant while keeping other parameters of the system unchanged, a numerical search has to be performed over a certain range of time, $t^*_{search}$. Moreover, to indicate the stability of the obtained advantage in uncertainty over time, we define $t^{*}_{range}$ as the total time duration for which the SQL is surpassed.

\begin{figure*}
    \centering
    \includegraphics[width=\linewidth]{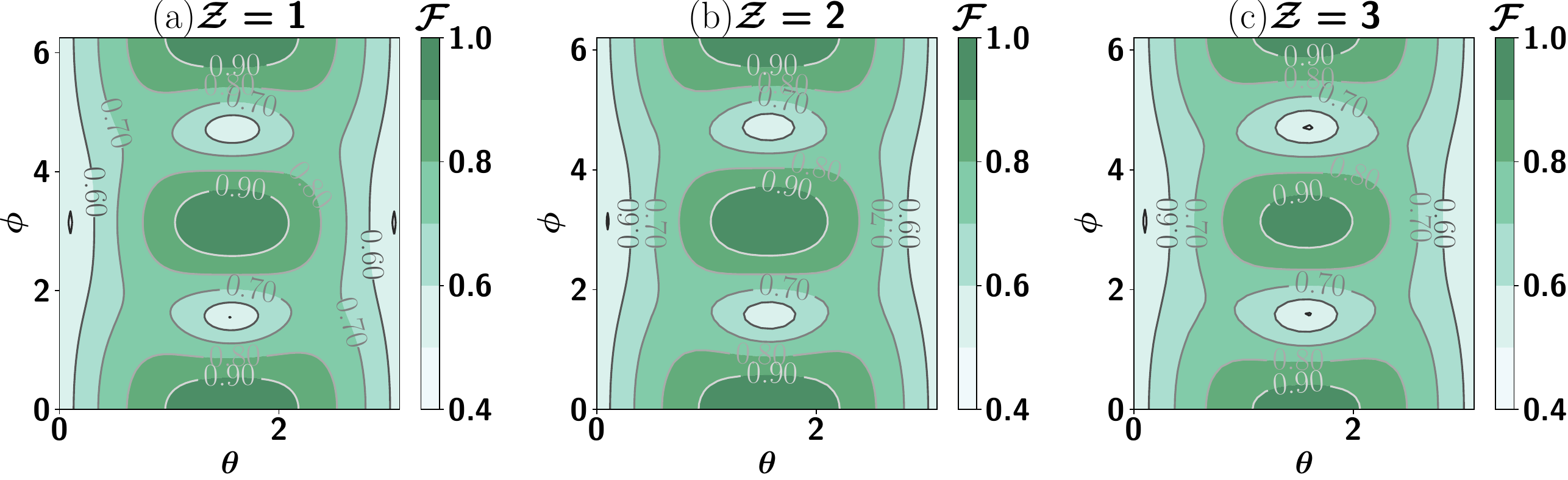}
    \caption{(Color Online.) Map plot of fidelity ($\mathcal{F}$) between the resulting state and the GHZ state (z-axis) with respect to the parameters involved in the measurement operators,  $\theta$ (x-axis) and $\phi$ (y-axis) for the long-range Hamiltonian. $\mathcal{Z}$ denotes the coordination number. The system size is $N = 4.$ Here $h^x = 0.05$, $\beta=10$, and $\alpha=1$. All the axes are dimensionless.}
    \label{fid_plot}
\end{figure*}

\subsection{Optimal measurement setting}
Since the estimation protocol involves arbitrary single qubit projective measurements to initiate the \emph{flipping mechanism} in the system, we optimize over all the $\theta$s and $\phi$s to find out the best possible choice of measurement settings to achieve the maximum fidelity with the GHZ state, thereby leading to minimum uncertainty. In particular, we observe that the fidelity with the $(N-1)$-qubit GHZ state, given in Eq. (\ref{eq:fidelity}), reaches its maximum value when $\theta = \pi/2$ and $\phi = n\pi$ where $ n = 0, 1, 2, \ldots$. It is well established that, in noiseless scenarios, the GHZ state provides Heisenberg scaling in parameter estimation protocols \cite{giovenneti2006}.  These specific values of $\theta$ and $\phi$ correspond to the optimal measurement basis, ${|\pm\rangle}$ where $|\pm\rangle = \frac{|0\rangle \pm |1\rangle}{\sqrt{2}}$ are the eigenvectors of $\sigma^{x}$ (see Fig. \ref{fid_plot} and Appendix \ref{app:secular}). Further, we perform another arbitrary rank-1 projective measurement on the same qubit on the evolved state, allowing us to extract information from the system by calculating the probability and hence the associated uncertainty in the estimation presented in Eq. (\ref{uncertainty}). We determine that the optimal measurement configuration, in this case, occurs for the basis,  $\theta^{\prime} = \pi/2$ and $\phi^\prime = n\pi/2$ where  $n = 1,3 \ldots$, i.e., for the basis, ${|\pm i \rangle}$ where $|\pm i\rangle   = \frac{|0\rangle \pm i |1\rangle}{\sqrt{2}}$, the eigenvectors of $\sigma^y$ (see Fig. \ref{fig:sigma_y meas} in Appendix. \ref{sec:optimalend} and the corresponding discussion). We emphasize that these settings are valid for other values of the coordination number (\(\mathcal{Z}\)) (see Figs. \ref{fid_plot} \textcolor{red}{(b,c)} and \ref{fig:sigma_y meas} \textcolor{red}{(b,c)}). Specifically, rigorous numerical simulations show that as \(\mathcal{Z}\) increases, the optimal measurement settings remain unchanged. Furthermore, our results indicate that the fidelity improves in the case of long-range Hamiltonians compared to \(\mathcal{Z} = 1\). The optimization in our work has been performed over all steps in the time evolution range (\(t^*_{range}\)). In the rest of the paper, we calculate $\delta \omega$, by considering these optimal measurement settings.

\subsection{Determining the range of $t^{*}_{search}$} 

We are interested to find out $t^*_{search}$ which finally leads to $t^*_{opt}$, the optimal time up to which the system should evolve according to the unitary $U_{t^*}$ such that $\mathcal{F}$ is maximum. Apart from checking for the optimal $\mathcal{F}$, the maximal magnetization density also suggests $t^*_{opt}$ since the optimal measurement on the edge qubit induces flipping (see Appendix: \ref{app:secular}) which implies that the measurement and $t^*_{opt}$ are interrelated. We notice that during the dynamics, there is a specific time at which $\frac{M^{z}}{N}$ reaches its maximum value for certain system parameters, indicating that all the spins except the spin in which measurement is performed are flipped in the same direction. Moreover, the state of the system at that instant of time shares maximum fidelity with the GHZ state. It turns out that the time at which $\frac{M^z}{N}$ is maximized can indicate the range of $t^*$ from which the optimal time $t^{*}_{opt}$ can be found for achieving the minimum uncertainty in determining the probe field, as illustrated in Table. \ref{tbl:bins} \cite{lee05, longrange_spin_wave_Khitrin_2006}. 

\begin{figure}
 \centering \includegraphics[width=\linewidth]{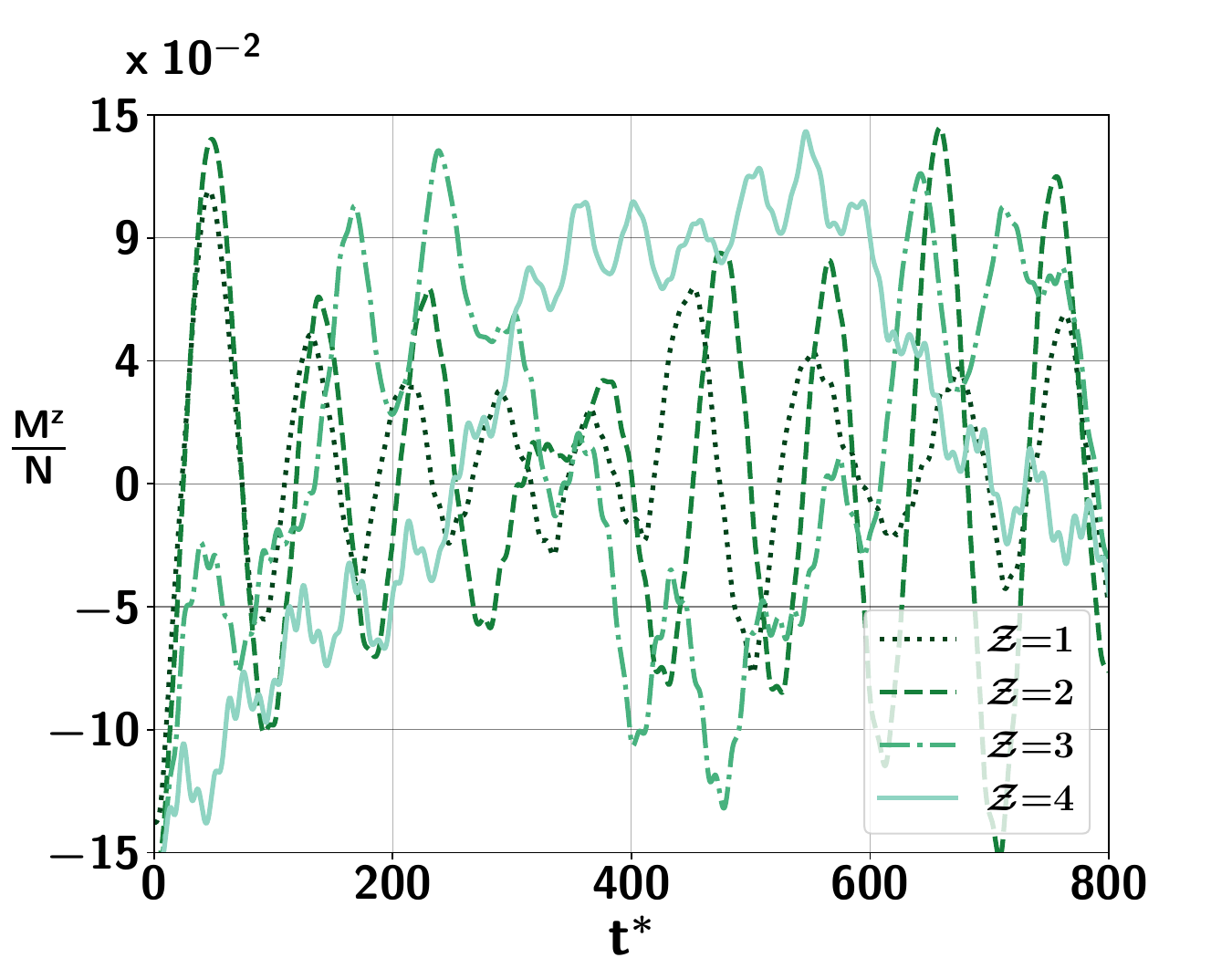}
 \caption{(Color Online.) Behavior of magnetization density, $\frac{M^{z}}{N}$(ordinate) with respect to time duration $t^{*}$(abscissa). Different lines represent different values of the coordination number, $\mathcal{Z}$. Here $N = 10$ and $h^x = 0.25$. All other specifications are same as in Fig. \ref{fid_plot}.
 Both the axes are dimensionless.} \label{mag_plot}
\end{figure}

Precisely, magnetization density exhibits a quasi-periodic nature with respect to time. As time increases, $\frac{M^{z}}{N}$ initially increases and then reaches its maximum value (see Fig. \ref{mag_plot}). Furthermore, with the increasing coordination number, the spread of magnetization also increases. Therefore, we fix the value of $t^{*}_{search}$ as double the value of $t^{*}_{opt}$, such that the maximum value of magnetization (i.e., minimum uncertainty) can be unambiguously determined. It is important to notice that as the value of the transverse magnetic field increases,  $t^*_{search}$ increases to accommodate spin flipping for longer times. On the other hand, the corresponding $\frac{M^{z}}{N}$ also decreases. Thus, we observe that it is optimal to consider the value of $h^x$ to be around $0.25$, irrespective of the coupling strength and the range of interactions. We will again address the desirable strength of the magnetic field of the sensor in detail.

\begin{table}

\begin{center}

\begin{tabular}{|c|c|c|} 
\hline
\textbf{$\mathcal{Z}$} & \textbf{$t^*_{opt}(\frac{M^{z}}{N})$} & \textbf{$t^*_{opt}(\delta\omega)$ } \\
\hline
$1$ & $46.1$   & $44.7$   \\
\hline
$2$ & $48.3$   & $48.0$     \\
\hline
$3$ & $238.4$  & $239.05$     \\
\hline
$4$ & $546.2$  & $548.25$  \\
\hline
\end{tabular}
\caption{Optimal times for achieving maximum magnetization density, \(t^*_{opt}(\frac{M^{z}}{N})\) and for  minimum uncertainty,  $t^*_{opt}(\delta\omega)$ for long-range Ising model with \(\alpha = 1\) for different values of \(\mathcal{Z}\). Here $N = 10$.}
\label{tbl:bins}
\end{center}
\end{table}

\subsection{Coordination number vs fall-off rate  in a quantum sensor}

Equipped with the optimal measurement settings and $t^{*}_{opt}$, we are now ready to investigate the role of coordination number and the range of interaction on the minimum uncertainty obtained in measuring the magnetic field. Specifically, we now illustrate that both the range of interaction and the coordination number are interconnected to estimate \(\omega\).

The long-range Ising model possesses different phases characterized by the correlation length between different pairs of spins and parameterized by $\alpha$, namely, non-local $(0 \le \alpha \le 1)$, quasi-local $(1 < \alpha < 2)$ and local $(\alpha \ge 2)$. Note that in physical systems like trapped ions, such Hamiltonian with variable-range interactions following power law naturally appears \cite{vodola2015longrangephases}. 

{\it Local regime.} We report that by preparing the system in the local regime, it is possible to beat the SQL for all coordination number $\mathcal{Z}$, as shown in Fig. \ref{diffalpha_plot} with $\alpha = 3.0$  although the SQL can be surpassed for a limited range of $t^*$, referred to a $t^{*}_{range}$ which can be experimentally challenging. 

\begin{figure}
 \centering \includegraphics[width=\linewidth]{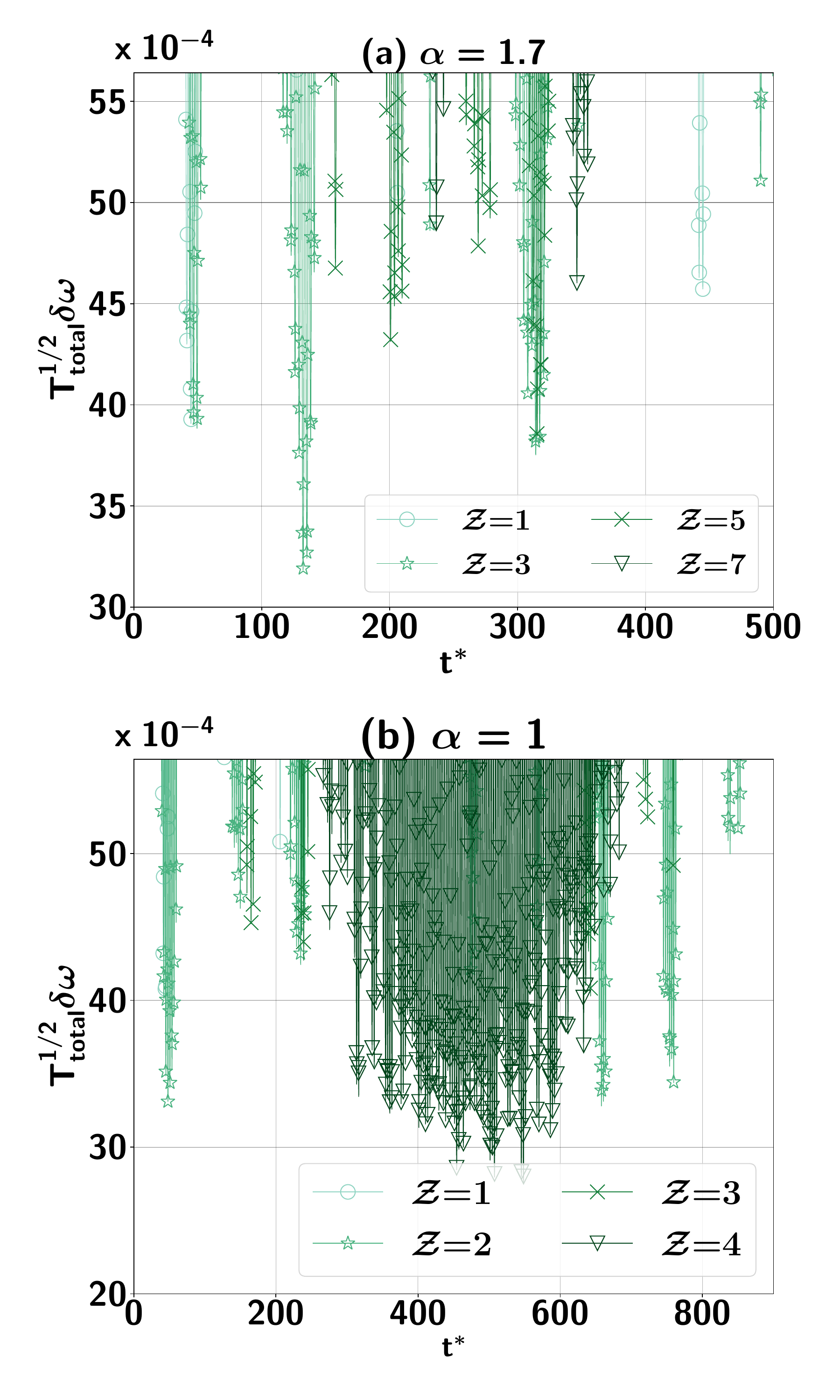}
 \caption{(Color Online.)  $T^{1/2}_{total}\delta\omega$ (ordinate)  vs the time interval in which the system evolves initially and in the final step, $t^{*}$ (abscissa) of the long-range transverse Ising chain as a sensor.  The upper and lower panels correspond to two different values of the power law decays, (a) \(\alpha =1.7\) and (b) \(\alpha =1\)  respectively. Here $\mathcal{Z}$ is the coordination number of the Ising Hamiltonian. The system size is $N=10$. The upper limits on the vertical axes represent the SQL. The other parameters involved in the sensing protocols are taken to be  \(h^x =0.25\), $t_{int}=1000\pi$, $\omega = 10^{-6}$, \(\alpha =1\), and \(\beta =10\).   Both axes are dimensionless.} 
\label{delw_plot}
\end{figure}

\begin{figure}
 \centering \includegraphics[scale=0.4]{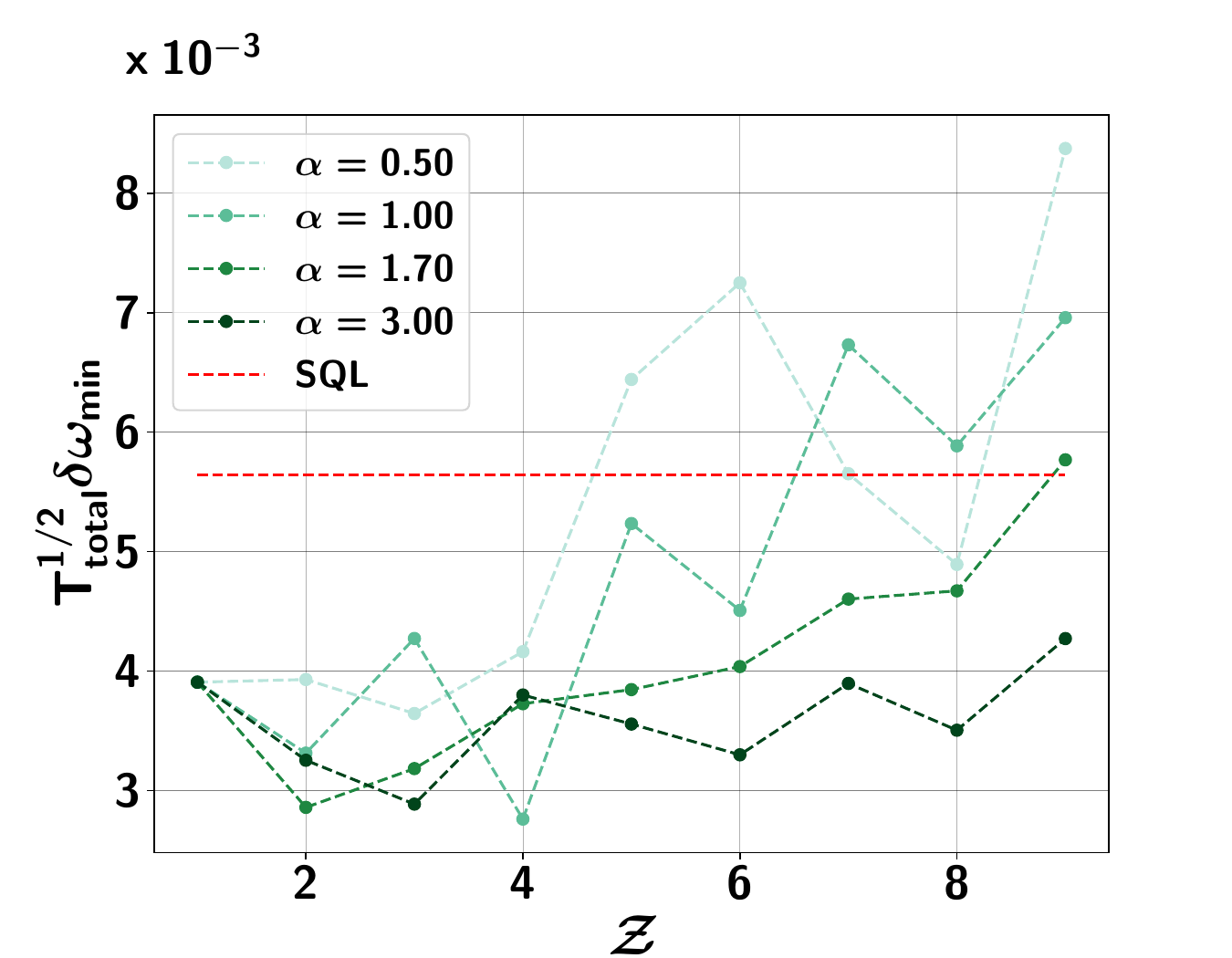}
 \caption{(Color Online.) {\bf Role of range of interactions on sensors.} Here $T^{1/2}_{total}\delta\omega_{min}$ against the coordination numbers $\mathcal{Z}$ (x-axis) for different fall-off rates ($\alpha$) of the interaction strength. Note that the sensor with \(\alpha \geq 1.7\) always beats the SQL for all values of \(\mathcal{Z}\). 
 All other specifications are same as in Fig. \ref{delw_plot}. 
 Both axes are dimensionless.} 
 \label{diffalpha_plot}
\end{figure}

{\it Non-local phase.} On the other hand, in the non-local regime, although $t^{*}_{range}$ increases significantly, indicating the experimental feasibility and the stability of the system, there exists an upper bound on the interaction range, such as $\mathcal{Z} = 6$ for $N = 10$ and \(\alpha =1\) (see Fig. \ref{delw_plot} (b)), beyond which the system cannot cross the SQL.  We find that $\delta\omega$ shows an oscillatory behavior with respect to the time of the evolution for various coordination numbers ($\mathcal{Z}$) as depicted in  Fig. \ref{delw_plot}. In other words,  when $\alpha \leq 1$, we observe that there exists a critical value of $\mathcal{Z}$ up to which the SQL limit is surpassed. For example, a strongly coupled model Hamiltonian with coordination numbers, $\mathcal{Z}=7, 8$, and $9$ for $N = 10$ cannot beat the SQL as depicted in Fig. \ref{diffalpha_plot}. This illustrates that while increasing the interaction range may enhance the sensitivity, it is not universally effective in building desired quantum sensors. However, from the experimental point of view, we find that the spread of uncertainty oscillations below the SQL increases with time, which increases with higher coordination numbers, up to a certain cut-off of $\mathcal{Z}$. This leads to flexibility in measurements and evolution time. 

{\it Gain in sensitivity in the quasi-local domain.} The interesting trade-off between $t^*$ and the minimum value of $\delta \omega$ emerges when the initial state is prepared in the quasi-local regime having $1 < \alpha < 2$. In particular, in this domain, the sensor can always go below the SQL for all values of $\mathcal{Z}$ (see Fig. \ref{diffalpha_plot} (a)), and the range of $t^*$ is moderately spread, as shown in Fig. \ref{delw_plot}  for $\alpha = 1.7$. With the decrease of $\alpha \lesssim 1.7$, there exists a critical value of $\mathcal{Z}$ above which the SQL cannot be exceeded.
This behavior is illustrated in Fig. \ref{diffalpha_plot}, where we plot $\delta\omega_{min}$ against the coordination numbers $\mathcal{Z}$ for different fall-off rates ($\alpha$) of the interaction strength. The study reveals that the sensor consistently outperforms the SQL for all $\mathcal{Z}$ when $\alpha \geq 1.7$. However, for $\alpha<1.7$, the sensor only surpasses the SQL up to a certain coordination number $\mathcal{Z}$.

It illustrates that minimum uncertainty can also be used to separate non-local from local regimes of the Ising Hamiltonian. We know that such separation is possible via the correlation length.  Specifically, if the sensitivity cannot exceed the SQL for all $\mathcal{Z}$, we can surely infer that the sensor may involve a long-range  Hamiltonian with \(\alpha \leq 1\) which for no \(\mathcal{Z}\) simply implies the sensor Hamiltonian  to be \(\alpha >2\). The latter observation remains true for the quasi-local Hamiltonian too with \(\alpha \lesssim 2\). The entire analysis reveals that  quantum sensors built with the long-range Hamiltonian having moderate $\alpha$ and interactions possess all the desired properties of sensors like low uncertainty and high $t^*_{range}$, which are preferable for quantum sensing. With the increase in fall-off rate $\alpha$, the advantage over the SQL becomes more pronounced, and this advantage is sustained over a wider range of interactions. This is due to the fact that, as $\alpha$ increases, the transverse magnetic field eventually reaches a value where it exceeds the next nearest neighbor interaction strength while remaining lower than the nearest neighbor interaction strength \cite{longrange_spin_wave_Khitrin_2006}. This condition triggers a flipping sequence, resulting in the generation of a polarization wave (see Appendix \ref{app:secular}). This flipping sequence allows the system to attain a state after evolution that closely resembles the GHZ state, providing minimum uncertainty in the estimation scenario. We will also show below that to extract its advantage, one is also required to suitably tune the external magnetic field. 

\begin{figure}
 \centering \includegraphics[width=\linewidth]{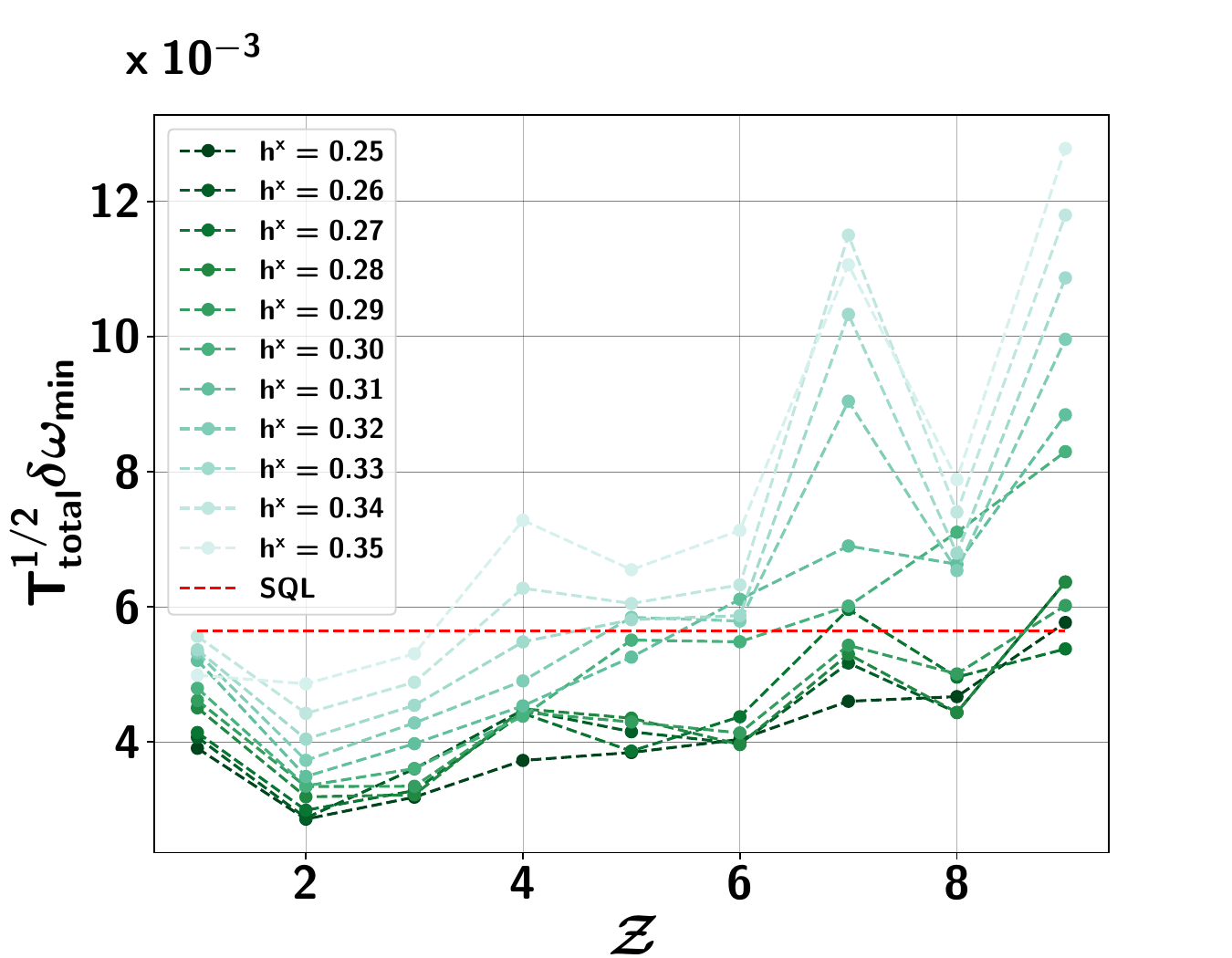}
 \caption{(Color Online.) {\bf Trends of minimum uncertainty $T^{1/2}_{total}\delta\omega_{min}$ (y-axis) with  $\mathcal{Z}$ (x-axis) for different $h^x$ values.} 
   All other specifications except \(\alpha\), which is chosen to be \(1.7\), are the same as in Fig. \ref{delw_plot}. 
 Both axes are dimensionless.
} \label{alpha1.7_plot}
\end{figure}

\subsubsection{Effect of external magnetic field}
Keeping the system in the quasi-local phase, increasing the value of the magnetic field of the sensor Hamiltonian provides a detrimental effect on the estimated uncertainty of the probe field. It also manifests that, to obtain a quantum advantage in sensors, the interaction between spins is required, which, in principle, can create multipartite entanglement.  The advantage over the SQL becomes less pronounced after reaching a certain value of $h^x$ as demonstrated in Fig. \ref{alpha1.7_plot}. The lowest uncertainty value is consistently observed for $\mathcal{Z}=2$ across all magnetic field strengths. Further, note that after a certain $h^x$ value, the advantage stays only up to $\mathcal{Z}=3$. In addition, the $t^{*}_{opt}$ value reduces significantly for higher $h^x$. This behavior is expected because, although for large $h^x$ values, the condition describing the propagation of the stimulated polarization wave is satisfied, the spin dynamics required for generating the GHZ state gets disrupted because the resonance condition is spoiled by increasing interactions with next-nearest neighbors, thus resulting in higher uncertainty values (see Appendix \ref{app:secular}).

\subsubsection{High temperature can overcome SQL with the aid of long-range interactions}

Upto now, the entire analysis has been performed with the initial state being prepared at a moderate temperature. However, the presence of thermal noise is an inherent factor that can not be avoided in experimental settings. Therefore, it is crucial to investigate whether the aforementioned advantages 
persist even when the system interacts with a thermal bath at high temperatures. 

 \begin{figure}
 \centering \includegraphics[width=\linewidth]{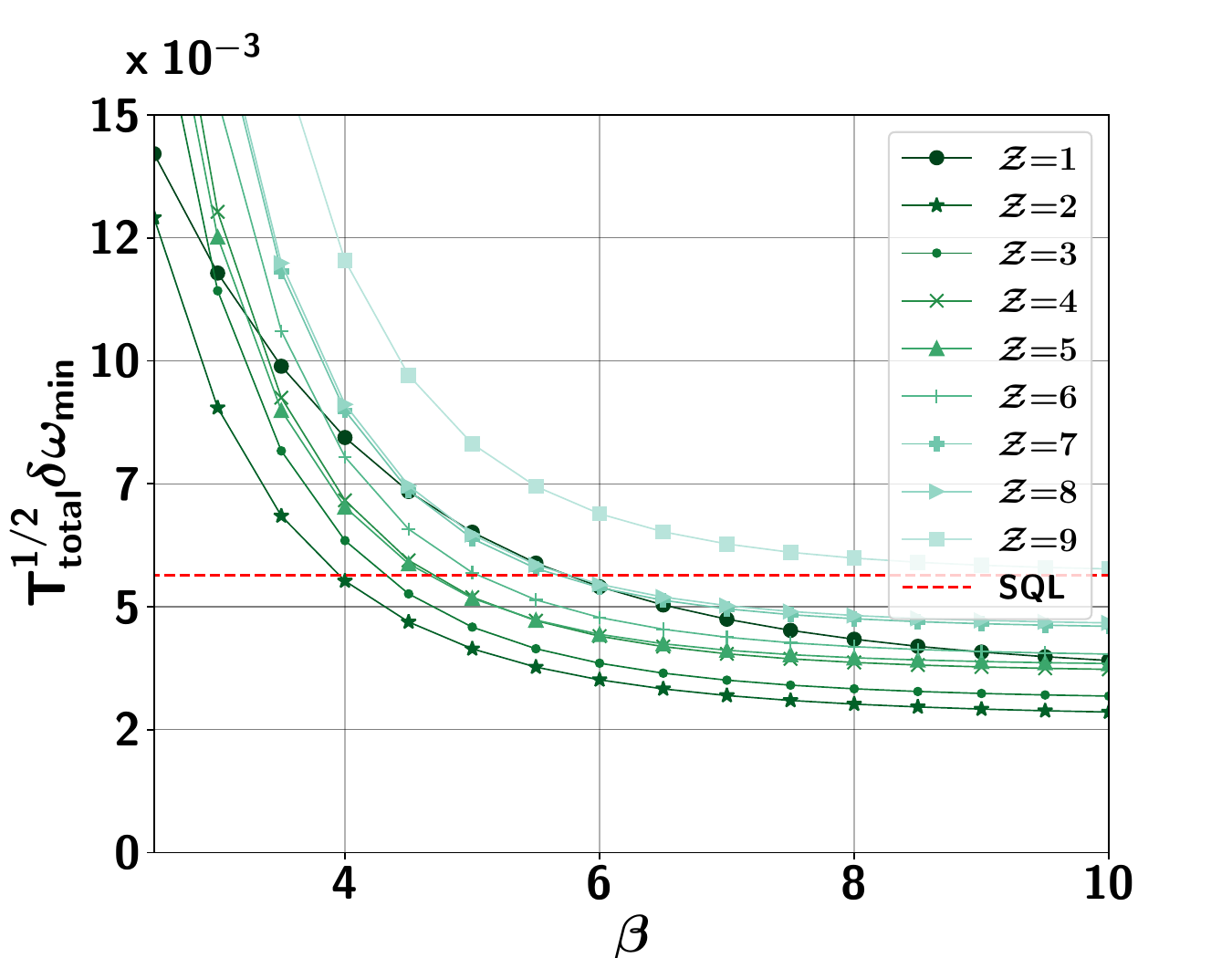}
 \caption{(Color Online.) {\bf Advantage of sensors with long-range Hamiltonian in presence of high temperature.} Here $T^{1/2}_{total}\delta\omega_{min}$ (vertical axis) is plotted with $\beta$ (horizontal axis) for different coordination numbers $\mathcal{Z}$. The red dashed line shows the SQL. Here \(\alpha =1.7\). All parameters used here are the same as in Fig. \ref{delw_plot}. Both axes are dimensionless. }
 \label{beta_plot}
\end{figure}
Counterintuitively, we find that by incorporating long-range interactions, it is possible to surpass thermal fluctuations and beat the SQL for higher temperatures where the sensor built with nearest-neighbor interactions fails. We define a critical value of inverse temperature, $\beta_{critical}^{\mathcal{Z}}$,  up to which quantum advantage persists for a given range of interaction. We report that
\begin{equation}
    \beta_{critical}^{\mathcal{Z} = 1}> \beta_{critical}^{\mathcal{Z} > 1},
\end{equation}
up to certain values of $\mathcal{Z}$ for a fixed value of $\alpha$. In particular, when $1 < \alpha < 2$, there exist values of $\mathcal{Z}$ for which  $\delta \omega_{\min}$ is obtained below the SQL for certain low $\beta$ values which cannot be reached with the nearest neighbor model. Specifically, when $\alpha = 1.7$ and $\mathcal{Z} = 2$, we find that uncertainty goes below the SQL with $\beta \approx 4$ which cannot occur with $\mathcal{Z} = 1$ (see Fig. \ref{beta_plot}).  This clearly indicates the advantage of using a long-range Hamiltonian as a quantum sensor in the presence of high temperatures. Notably, as $\beta$ increases, the minimum uncertainty value exhibits a slight decrement before rapidly reaching a saturation point. Beyond this point, further increments in $\beta$ do not yield significant improvements in reducing uncertainty. This observation opens up new possibilities for the development of sensors that can outperform the SQL, even in the high-temperature limit.


\subsubsection{System-size dependence in the parameter estimation}

\begin{figure}
 \centering \includegraphics[width=\linewidth]{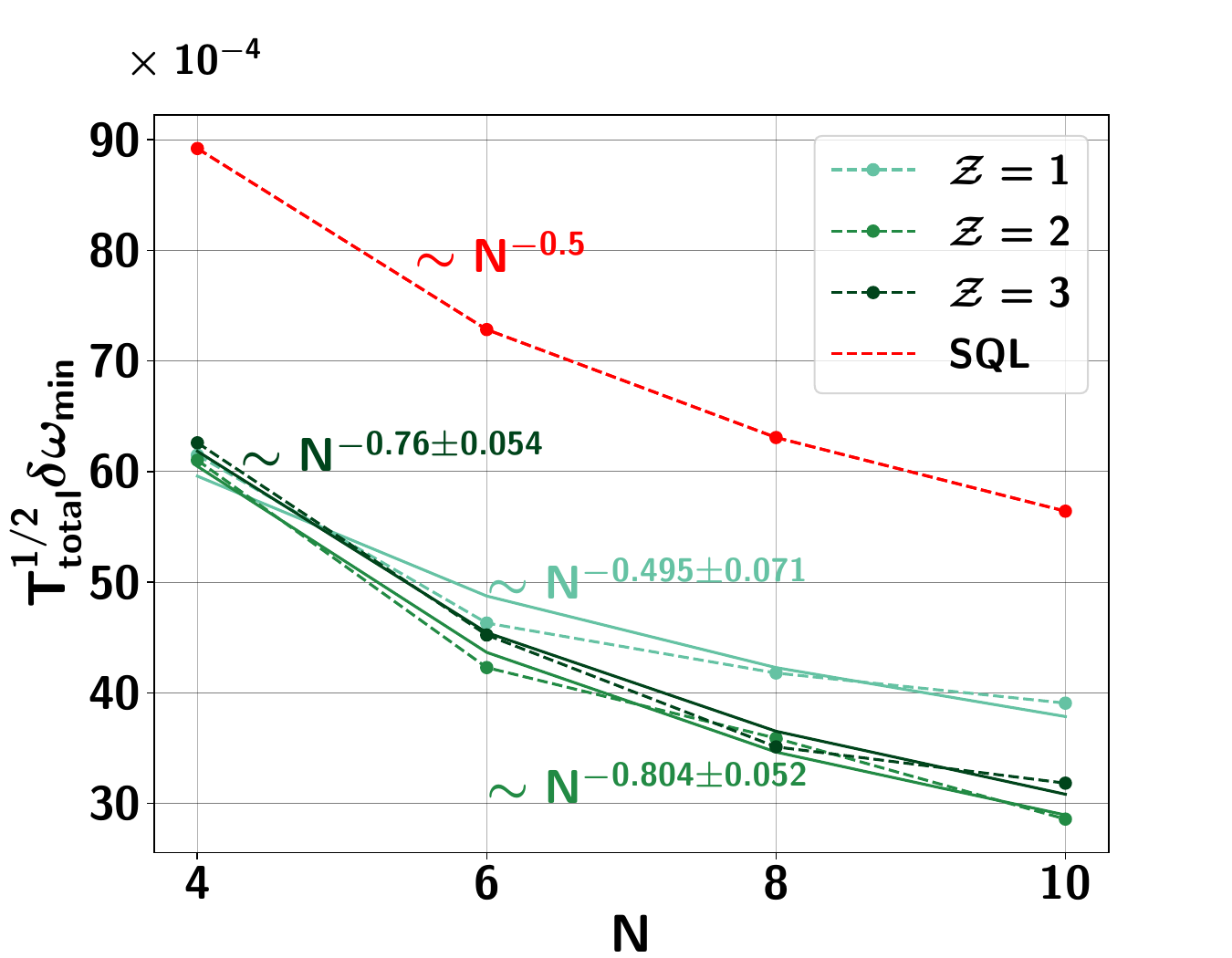}
 \caption{(Color Online.) Variation of uncertainty, $T^{1/2}_{total}\delta\omega_{min}$ (ordinate) with system-size \(N\) (abscissa). $\mathcal{Z}$ is the coordination number of the Ising Hamiltonian. The red dashed line shows the standard quantum limit (SQL). Here other parameters involved in the sensing protocols are taken to be \(\alpha =1.7\), $t_{int}=1000\pi$, $\omega = 10^{-6}$, and \(\beta =10\). For \(N\) values of \(4, 6, 8\), and \(10\), the corresponding \(h\) values are \(0.05, 0.15, 0.22\), and \(0.25\). The scaling for each coordination number and the corresponding SQL are $0.495\pm0.071,~0.80\pm0.052, 0.76\pm0.054$, and $0.5$ respectively, written in the same color as uncertainty. Both the axes are dimensionless.}
\label{delw_vs_N}
\end{figure}

It is well established that the total number of particles in a many-body quantum sensor is a resource for quantum sensing \cite{multiparameter_caves_2007, quantum_control_campo_2022, abolfazal_sequential_2022}. We show that with the increase of the system-size, $N$, uncertainty in the estimation of $\omega$ reduces considerably, thereby indicating the beneficial role of large system-size (see Fig. \ref{delw_vs_N}). We observe that the introduction of next-nearest neighbor interaction along with the nearest-neighbor one enhances the estimation accuracy compared to the spin chain with $\mathcal{Z} = 1$ for all $N$.
Our protocol significantly magnifies the precision beyond the standard quantum limit (SQL). Furthermore, we report (in Fig. \ref{delw_vs_N}) that the uncertainty in the estimation of $\omega$ scales as $N^{-0.8}$, thereby showing an advantage over the SQL( approximately equal to $N^{-0.5}$).
\begin{figure}
 \centering \includegraphics[width=\linewidth]{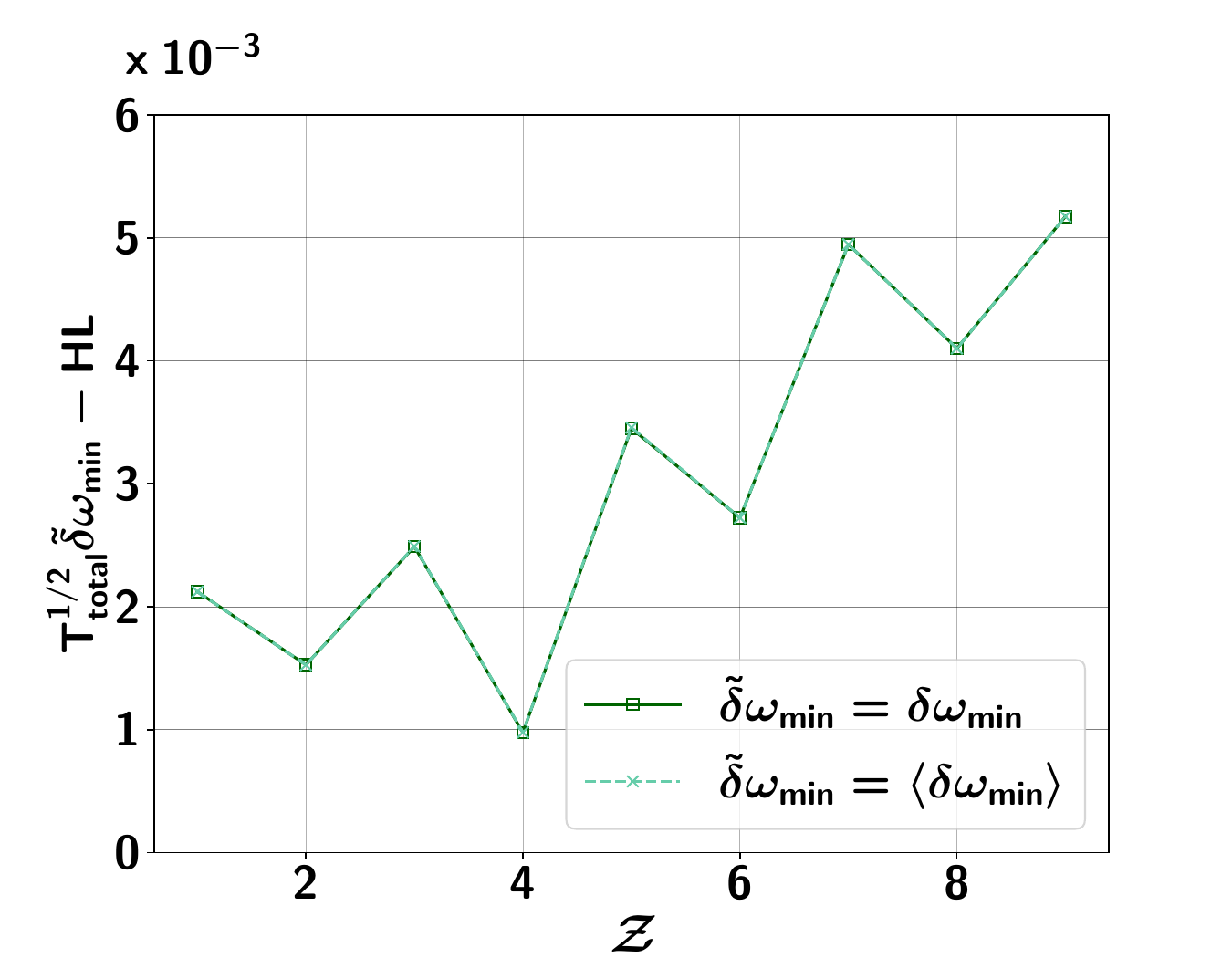}
 \caption{(Color Online.) {\bf Robustness of sensor against impurities.} Deviation of quenched averaged minimum uncertainty from the Heisenberg limit (HL), i.e., $T^{1/2}_{total}\tilde{ \delta}\omega_{min} - \text{HL}$ (ordinate) against $\mathcal{Z}$(abscissa). Let $\langle {\delta}\omega_{min} \rangle$ represent the average value of minimum uncertainty in the presence of disorder (dashed cross), and $\delta \omega_{min}$ represent the same in the absence of disorder (solid square). The Heisenberg limit for system-size $N=10$ is $1/(N\sqrt{t_{int}})$ with $t_{int}=1000\pi$. Mean value $\bar{h}_x$ and standard deviation $\sigma_G$ are taken to be $0.25$ and $0.0001$ respectively. Quenched averaging is performed over $500$ realizations. All the parameters used here are the same as in Fig. \ref{delw_plot}. Both the axes are dimensionless. } 
 \label{disorder_plot}
\end{figure}

\section{Robustness of sensors with long-range interactions against imperfections}
\label{sec:imperfections}

Imperfections can always disturb the performance of the quantum protocols, which is also the case for quantum sensors. Here we concentrate on two such situations -- (1) imperfections enter during the preparation of the initial state which implies the introduction of randomness either in the strength of the magnetic field or in the coupling constants; and (2) when the system is in contact with the bath, the sensing protocols get affected. We will show that the minimum uncertainty is robust against both the defects.

\subsection{Disorder-affected sensors}
\label{sec:disorder}

The amount of uncertainty involved in the parameter estimation protocol through the Cram{\'e}r-Rao bound \cite{cramer1946mathematical} clearly states that the error can be minimized by repeating the measurements a large number of times as seen in Eq. (\ref{uncertainty}). Therefore, in the course of repeating the experiment several times, erroneous realization of the probe, the measurement device, and most importantly, the inherent impurities in the sensor can adversely affect the performance of the sensor. Precisely, the practical implementation of the given protocol is likely to be affected by various sources of noise and disorder, such as environmental perturbations or imperfections in the experimental setup \cite{PhysRevLett.95.170410,PhysRevLett.98.130404,PhysRevLett.102.055301,Shapiro_2012,PhysRevA.72.063616}. 
Let us concentrate the effect of impurities present in the system on the performance of the sensor. Here we consider the quenched type of disorder in which the relevant observation time is much shorter than the equilibration time of the disordered parameter of the system. In order to examine the robustness of the performance of our proposed sensor, we deal with a sensor Hamiltonian in which the strength of the magnetic field is chosen randomly from two types of distributions, namely, Gaussian and uniform distributions (see Appendix \ref{disorder}). 

Following the same prescription as described in the preceding section, we measure the quenched average uncertainty of the estimated parameter, i.e., $\langle \delta \omega_{\min} \rangle$ by averaging over $500$ realizations of $\delta\omega$, each of which is obtained by an individually chosen magnetic field at each site from either a Gaussian (G) or a uniform (U) distribution with a mean value of $\bar{h}^x$ and a certain standard deviation $\sigma_{\xi}$ (with $\xi$ = G, U) (see Appendix \ref{disorder} for a description of both the distributions), representing the strength of the disorder. 
We check the convergence of the quenched averaged quantity up to six decimal places.

We observe that the performance of the suggested protocol is not significantly altered by either Gaussian or uniform disorder, in the presence of a moderate strength of disorder, thus implying the robustness of the proposed model against disorder. In particular, the uncertainty of the estimated parameter $\omega$ remains preserved even in the presence of disorder, i.e., it goes below the SQL as depicted in Fig. \ref{disorder_plot} provided the disorder strength is below some critical value, and the strength of magnetic field and other parameters are suitably tuned. These findings demonstrate the potential of our protocol for practical applications in real-world scenarios where the presence of disorder is inevitable.

Furthermore, we find that the maximum standard deviation at which the uncertainty exceeds the SQL value remains almost unaffected by the coordination number. This implies that the largest level of disorder, quantified by the standard deviation above which no advantage over the SQL is obtained, denoted by $\sigma_{max}$, remains relatively constant across all coordination numbers. The value of $\sigma_{max}$ exhibits minimal variation regardless of the specific coordination number under consideration, which indicates the robustness of the probe against disorder.

\subsection{Environmental effects on parameter estimation}
\label{decoherence}
Let us move to a situation when the sensor is in contact with the bath. It was shown that the Ramsey protocol for quantum sensing \cite{huelga, lee2002} does not provide the metrological advantage when the Markovian  noisy environment is present \cite{huelga}, i.e., the correlation time of the noise is much shorter than the interaction time between the probe and the target field. If the system undergoes evolution under uncorrelated noise so that the corresponding entanglement resource decays rapidly, the separable states and entangled states provide the same uncertainty in estimating a parameter \cite{huelga}. In Ref. \cite{matsuzaki2011}, it was shown that under a similar dephasing noise, precise engineering over noise correlation time and characteristic time of the system can achieve better estimation. More specifically, if the time of dephasing, $(T_{dph})$, is comparable to or longer than the exposure time $(t_{int})$ of the target field, the evolution would be non-Markovian and hence quantum advantage can be regained  \cite{matsuzaki2011}  see Appendix. \ref{app:time_inhomo}). With the sensor having long-range interactions, we then search for the appropriate interaction time as $t^{open}_{int}$, a value that can overcome the SQL.

We perform the evolution in the presence of the environment when the probe is interacting with the target field. The corresponding Lindbladian equation is given as 
\begin{equation}
\frac{d}{d t} \rho(t)=-i\left[H_{int}+H_{probe}, \rho(t)\right]-\frac{t}{2 T_{dph}^2} \sum_{i=1}^N\left[\sigma_i^z,\left[\sigma_i^z, \rho(t)\right]\right]. 
\end{equation}
Under the zeno-like measurement assumption of $T_{dph}>>t^{open}_{int}$, the above equation is analytically solvable as 
\begin{equation} \label{noiseeq}
\rho\left(t^{open}_{\mathrm{int}}\right)=\mathcal{M}_1\left(\mathcal{M}_2\left(\cdots \mathcal{M}_N\left(\hat{\rho}_I(t^{open}_{int})\right) \cdots\right)\right),
\end{equation}
such that $ \mathcal{M}_n(\rho)= p' \rho+ (1-p') \sigma_n^z \rho \sigma_n^z$, where $p' = \frac{1+e^{-(t^{open}_{\mathrm{int}} / T_{dph})^2}}{2}$ and $ \rho_I(t^{open}_{int})=e^{-i\left(H_{int}+H_{probe}\right) t^{open}_{\mathrm{int}}} \rho(t^*) e^{i\left(H_{int}+H_{probe}\right) t^{open}_{\mathrm{int}}}$. After the final step of $U_{t^*}$, we perform the measurement to find the probability. It is important to note that the corresponding $\left|\partial p/\partial \omega\right|$ depends on the value of $t^{open}_{int}$ and therefore, the optimization is performed to obtain  $t^{open}_{int}$ such that the uncertainty in Eq. (\ref{uncertainty}) is minimized  \cite{matsuzaki2011, chin2012, tatsuta_decoherence_2019}. Notice that the optimization of the time span, \(t_{int}\), during which the probe interacts with the sensor is not required in the noiseless scenario.

\begin{figure}
\centering \includegraphics[width=\linewidth]{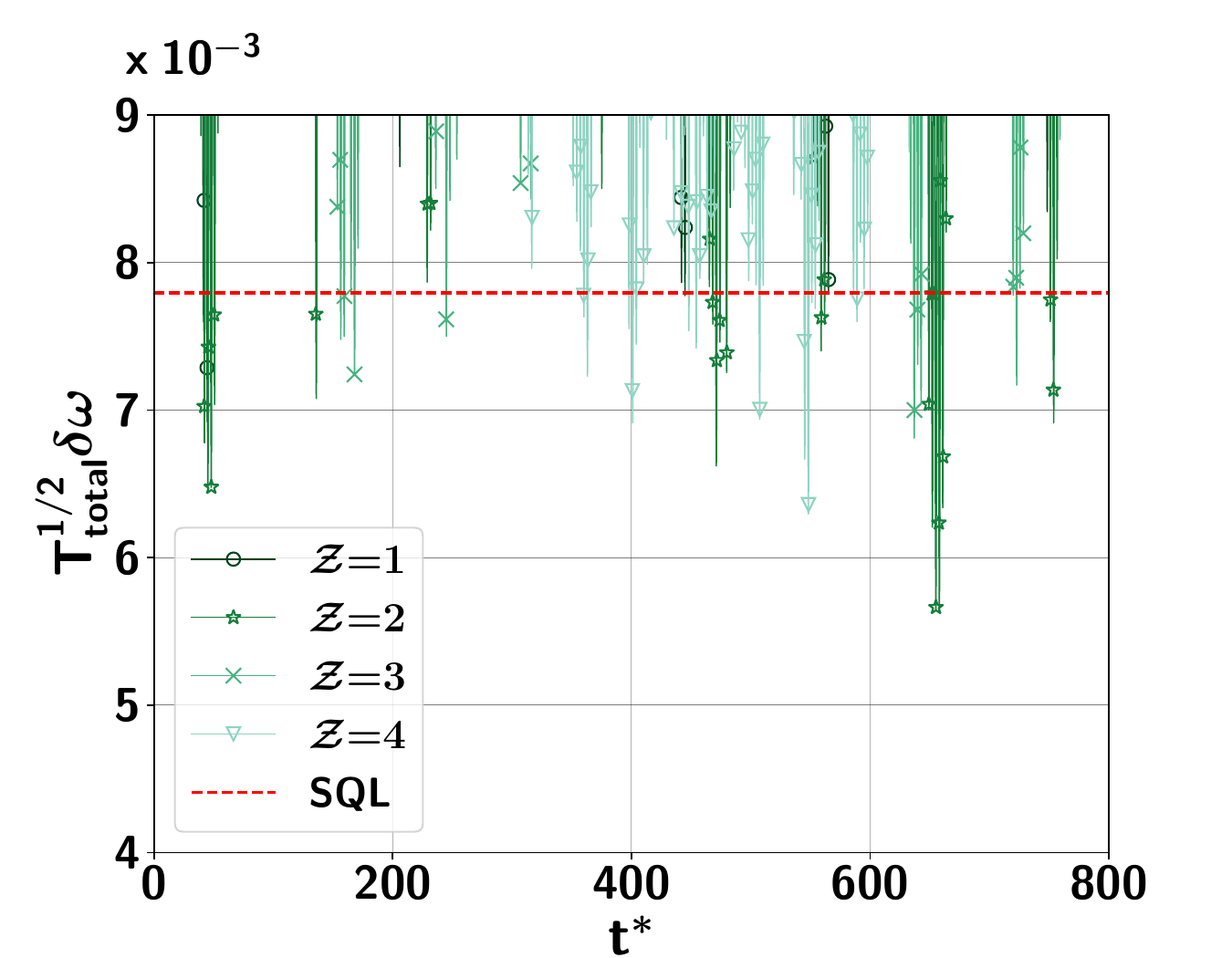}
\caption{ (Color Online.) {\bf Non-Markovian-type dephasing noise on sensors.} $T^{1/2}_{total}\delta\omega$(y-axis) with respect to time duration $t^{*}$(x-axis) in presence of dephasing noise, described in . 
Here $h^x = 0.25$,
$t_{int}^{open}=1646$, $T_{dph}=10^{4}$, $\beta=10$,  and $\alpha=1$. All the axes are dimensionless.} 
\label{dp_fig}
\end{figure}

The numerical simulation reveals that the quantum Ising model with long-range interactions outperforms the SQL against time-inhomogeneous dephasing noise. The enhanced sensitivity in the case of the long-range Hamiltonian is more pronounced than that of the system Hamiltonian with nearest-neighbor interactions (comparing \(\delta \omega\) with \(\mathcal{Z} =1\) and \(\mathcal{Z} >1 \)  as shown in Fig. \ref{dp_fig}). However, the benefit in sensitivity persists only up to coordination number six when $\alpha = 1$. Furthermore, in the presence of time-inhomogeneous dephasing noise, we note that the improvement over the SQL is smaller compared to the noiseless case (see Figs. \ref{delw_plot} and \ref{dp_fig}).

\section{Conclusion}
\label{conclusion}

Quantum sensors can improve their efficiency in estimating system parameters by utilizing entangled states. The strength of a magnetic field, for example, is estimated by creating interactions between a quantum spin chain and the probe field. 

We explored the effects of long-range interactions in the spin chain used as a quantum sensor on the estimation of the magnetic field. The sensing protocol depends on several key factors. One is the range of interactions, which divides the spin chain into three distinct phases: nonlocal, quasilocal, and local. Other important parameters include the coordination number, the temperature of the initial state, and the time interval for system evolution and its interaction with the probe. The methodology used in this work includes measurements of a single qubit both at the beginning and end. 

We found that the quantum sensor based on the transverse Ising chain with long-range interactions does not always exhibit quantum advantage by exceeding the standard quantum limit (SQL) for all coordination numbers, despite the fact that the situation is much better for the quasi-long-range Hamiltonian in terms of precision and time period. Additionally, we discovered that the precision obtained with the inclusion of next-nearest neighbor interactions outperforms the model with nearest-neighbor interactions. Further, the scaling of a sensor involving many-body Hamiltonian with nearest- and next nearest-neighbor interactions improves with the system-size compared to the SQL, thereby establishing its quantum advantage.

We demonstrated that the quantum advantage may be acquired even when the initial state used for estimating the probe can be prepared in a canonical equilibrium state with moderate to high temperature,  provided the system possesses quasi-long range interactions. In the presence of disorder in the transverse magnetic field of the sensor, the quenched average minimum uncertainty remains almost unaltered provided the disorder strength is moderate. We also showed that when the system interacts with non-Markovian-type dephasing noise, the beneficial role of the long-range model still survives. 

In summary, our findings emphasize the potential of long-range interacting quantum systems with decay rates, demonstrating their enhanced metrological performance, robustness against imperfections, and superiority in the presence of environmental interactions.  These findings support the fabrication of quantum sensors in physical systems that cannot avoid long-range interactions.

\begin{acknowledgements}

We acknowledge the support from Interdisciplinary Cyber Physical Systems (ICPS) program of the Department of Science and Technology (DST), India, Grant No.: DST/ICPS/QuST/Theme- 1/2019/23. We acknowledge the use of \href{https://github.com/titaschanda/QIClib}{QIClib} -- a modern C++ library for general purpose quantum information processing and quantum computing (\url{https://titaschanda.github.io/QIClib}) and cluster computing facility at Harish-Chandra Research Institute. This research was supported in part by the ’INFOSYS scholarship for senior students’. LGCL is funded by the European Union. Views and opinions expressed are however those of the author(s) only and do not necessarily reflect those of the European Union or the European Commission. Neither the European Union nor the granting authority can be held responsible for them. This project has received funding from the European Union’s Horizon Europe research and innovation programme under grant agreement No 101080086 NeQST. This work was supported by the Provincia
Autonoma di Trento, and Q@TN, the joint lab between University of Trento, FBK—Fondazione Bruno Kessler,
INFN—National Institute for Nuclear Physics, and CNR—National Research Council.

\end{acknowledgements}
\appendix

\section{Secular Hamiltonian for long-range interacting systems: Condition for the generation of spin-polarization wave}
\label{app:secular}

\begin{figure*}
    \includegraphics[width=\linewidth]{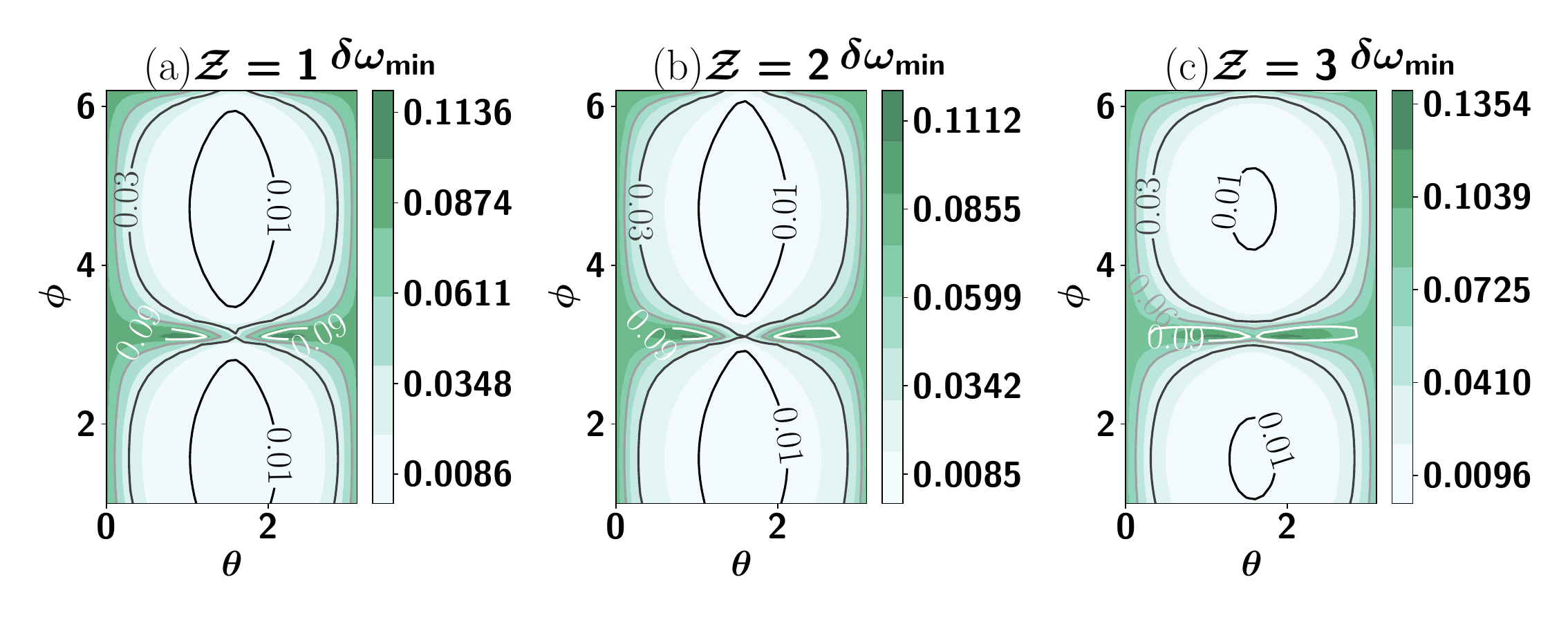}
    \caption{(Color Online.) Map plot of minimum uncertainty ($\delta \omega_{min}$) with respect to the parameters involved in the measurement operators,  $\theta$ (x-axis) and $\phi$ (y-axis) for the long-range Hamiltonian. $\mathcal{Z}$ is the coordination number of the long-range Hamiltonian. The system size is $N = 4.$ Other parameters in the protocol used here are $h^x = 0.05$, $\beta=10$, and $\alpha=1$. All the axes are dimensionless.}
    \label{fig:sigma_y meas}
\end{figure*}
The spin polarization wave can be generated using the Ising model with long-range interactions. In the work by Furman {\it et al.} \cite{longrange_spin_wave_Khitrin_2006}, it was shown that given an initial state, $|100\ldots0\rangle$ where $|0\rangle$ and $|1\rangle$ are the ground and excited states of the $\sigma^z$, the evolution governed by the unitary $U_{t^*}$ of the long-range Hamiltonian (in Eq. (\ref{H_tfi})) induces a flipping sequence in the spin chain, known as the domino dynamics. 

Let us explain the mechanism with a particular example of a spin chain of a total number of sites $N = 4$ which exhibits the role of $U_{t^*}$ in this scheme. Suppose the initial state of the system is $|1000\rangle$. It was shown that there exists an evolution operator, $U_{t^*}$, which generates a spin polarization wave such that the resultant state becomes $U_{t^*}|1000\rangle = |1110\rangle$. Note that the domino dynamics can take place in the reverse direction as well, i.e., if one starts with $|1110\rangle$, after the application of a unitary, the final state becomes $|1000\rangle$. Now, instead of starting with $|0000\rangle$, we prepare the initial state as $|+\rangle \otimes |000\rangle$ by performing a rank-one projective measurement on the boundary qubit. Precisely, we project the first spin in the subspace of $|+\rangle\langle+|$, where
$|+\rangle$ is the eigenvector of $\sigma^x$ Pauli matrix. Now the evolution by the aforementioned unitary produces the state $U_{t^*}(|+\rangle \otimes |000\rangle) = \frac{1} {\sqrt{2}}|0000\rangle+|1110\rangle$, i.e., $ \frac{1} {\sqrt{2}}(|000\rangle+|111\rangle)|0\rangle  = |GHZ\rangle \otimes |0\rangle$. It can now be used as a resource for the quantum sensing protocols.

After encoding the information of the target field in the sensor by interacting the system with the target field, via $U_{t_{int}}$ , if we again apply the unitary $U$ for the same interval of time $t^*$, following the same dynamics, the boundary spin becomes disentangled from the rest of the spin chain. In addition, all the information about the target field is then accumulated in that boundary spin. Consequently, in the readout session, we just require to perform a suitable local measurement on that qubit as opposed to the entire state which can be expensive for gathering information about the target field.

To obtain the effective (secular) Hamiltonian responsible for the generation of this spin polarization wave, the Hamiltonian is transformed into the interaction picture, and the time-dependent part in the interaction picture is neglected. In this context, all the spins are irradiated at resonant frequencies by perturbation term $H_{field} = h^x \sum _i \sigma^x$. In their study, they have proven that by considering the limit $J_{ij}\gg h^x$ for $\alpha=3$, the secular Hamiltonian can be expressed as follows:
\begin{equation} \label{H_secular}
H_{eff}=\frac{h^x}{2} \sum_{i=\mathcal{Z}+1}^{N-\mathcal{Z}} \sigma_i^x \; \prod_{j=1}^\mathcal{Z}\left(1-4 \sigma_{i+j}^z \sigma_{i-j}^z\right)+ H_{edge}.
\end{equation}
Here, $H_{edge}$ represents the Hamiltonian 
representing all the connections at the edge. 
However, as stated in Ref. \cite{longrange_spin_wave_Khitrin_2006}, describing the spin dynamics of the polarization wave using the Hamiltonian in Eq. (\ref{H_secular}) is not possible due to the interaction with next-nearest neighbors, which spoils the resonance condition. This issue can be addressed by assuming a moderate value of the transverse magnetic field $h^x$ that is greater than the next nearest-neighbor interaction strength but less than the interaction strength of the nearest neighbors. These conditions, combined with a weak resonant transverse magnetic field, lead to a flipping sequence in the spin chain, giving rise to the polarization wave, thereby leading to a GHZ state, responsible for quantum advantage in sensing.

\section{Optimal choice of the measurement basis at the end of the protocol}
\label{sec:optimalend}

To maximize the precision, thereby optimizing the performance of the sensor, it is crucial to choose a suitable measurement basis at the end of the protocol. Note first that the measurement basis derived from the eigenbasis of the symmetric logarithmic derivative of the encoded state is widely recognized as the optimal set of measurement operators that achieve saturation of the Cramér-Rao bound \cite{sld_saturation}. The reasons behind choosing the eigenvectors of $\sigma_{y}$ as our optimal measurement setting at the final step of the protocol are two-fold -- First, our protocol closely resembles the typical Ramsey measurement setup, where a GHZ state is created, and utilized as a resource. Given this similarity, it is intuitive to adopt the measurement used in the Ramsey setup which is precisely the measurement in the eigenbasis of $\sigma_{y}$. Second, rigorous numerical simulation (via optimizing the minimum uncertainty against arbitrary projective measurements) also reveals that the optimal measurement is indeed the eigenbasis of $\sigma_{y}$ as shown in Fig. \ref{fig:sigma_y meas}. 

\section{Gaussian and uniform disorder}
\label{disorder}
In  Gaussian disorder, the strength of the transverse magnetic field $h^x$ involved in the Hamiltonian in Eq. (\ref{eq:xy_model}) is chosen randomly from a Gaussian distribution. Hence, the probability density function of choosing \(h^x\) from this distribution can be expressed as
   \begin{equation}
    P(h^x) = \frac{1}{\sigma_G\sqrt{2\pi}} \exp \left(-\frac{(h^x - \bar{h}^x)^2}{2\sigma^2_G}\right),
    \end{equation}
where $\bar{h}^x$ is the mean value of the magnitude of the magnetic field in the transverse direction, and $\sigma_G$ is the standard deviation of the distribution \cite{PhysRevB.102.245125}. Here, the subscript G in the mean and standard deviation represents the Gaussian distribution. The quenched averaged value of the physical quantity, which is the minimum uncertainty here, is given by \(\langle \delta \omega \rangle = \int P(h^x) \delta \omega_{\min}(h^x) dh^x\). 

in the case of uniform disorder, $h^x$ is distributed as
    \begin{equation}
    P(h^x) = \begin{cases} \frac{1}{b-a} & a \leq h^x \leq b, \\
    \ 0 &\text{otherwise,} \end{cases}
    \end{equation}
    where $a$ and $b$ are the lower and upper bounds of the range of the uniform distribution. The mean and standard deviation of the uniform distribution are given by
    $\bar{h}^x = \frac{a+b}{2}$ \text{and,}
    $\sigma_U = \sqrt{\frac{(b-a)^2}{12}}$ respectively \cite{PhysRevLett.95.170401,PhysRevA.95.021601}. In a similar fashion, the quenched average quantity can be computed.  

\section{Time-inhomogeneous noise}
\label{app:time_inhomo}

The physical mechanism that allows for a quantum sensor to perform better than the shot noise limit is due to the presence of non-Markovian noise. IN particular, in the presence of time-correlated noise, it has been shown that the coherence of a given state does not decrease drastically as compared to the uncorrelated one \cite{matsuzaki2011}. The interaction with the environment can be written as $H_I=\lambda g(t) \sigma^z$, where the function $g(t)$ is the normalized Gaussian noise. This essentially means that the $\sigma^z$ operator acts on each qubit with a corresponding Gaussian probability in time $t$. Let us suppose $\rho_0$ is the initial state of the system. In the interaction picture, the evolution of $\rho_0$ can be written as
\begin{equation}
\begin{aligned}
\rho_I(t) &= \rho_0 + \sum_{n=1}^{\infty}(-i \lambda)^n \int_0^t d t_1 \int_0^{t_1} d t_2 \cdots \int_0^{t_{n-1}} d t_n \\
& \times\left[H_I\left(t_1\right),\left[H_I\left(t_2\right), \ldots,\left[H_I\left(t_n\right), \rho_0\right] \cdots\right]\right].
\end{aligned}
\end{equation}
The noise is said to be time-inhomogeneous in the sense of   
\begin{equation}
\overline{g\left(t_1\right) g\left(t_2\right)}=\frac{2}{\sqrt{\pi}} e^{-\left(\left|t_1-t_2\right|^2\right) / \tau_c^2},
\end{equation}
where $\tau_c$ indicates the amount of time over which the noise is correlated. The noise is considered to be Markovian if $t >> \tau_c$, where $t$ denotes the time of exposure of the given quantum system with the noise. On the other hand, if $t << \tau_c$, the interaction is considered to be non-Markovian. To understand why the non-Markovian noise is beneficial for quantum sensing, consider an initial state  $|+\rangle = \frac{|0\rangle + |1\rangle}{\sqrt{2}}$. In the former scenario, the off-diagonal terms follow $\langle 0| \rho_I(t)|1\rangle \simeq \frac{1}{2} e^{-4 \lambda^2 \tau_c t}$ while in the latter case,  $\langle 0| \rho_I(t)|1\rangle \simeq \frac{1}{2} e^{-(4 / \sqrt{\pi}) \lambda^2 t^2}$ (quadratic decay). This indicates that the system remains coherent throughout the exposure to the noise. In relation to quantum sensing, the behavior of the GHZ state  under such a noise is derived \cite{matsuzaki2011} as \begin{equation}
\begin{aligned}
\rho(t)= & \frac{1}{2}\left(\bigotimes_{l=1}^L|0\rangle_l\langle 0|\right)+\frac{1}{2} e^{i L \omega t-L \gamma(t) t}\left(\bigotimes_{l=1}^L|0\rangle_l\langle 1|\right) \\
& +\frac{1}{2} e^{-i L \omega t-L \gamma(t) t}\left(\bigotimes_{l=1}^L|1\rangle_l\langle 0|\right)+\frac{1}{2}\left(\bigotimes_{l=1}^L|1\rangle_l\langle 1|\right), 
\end{aligned}
\end{equation}   
where the decoherence of the single qubit is given by
$
\gamma(t)=\frac{4 \lambda^2 \tau_c^2\left(-1+e^{-\left(t^2 / \tau_c^2\right)}\right)}{\sqrt{\pi} t}+4 \lambda^2 \tau_c \operatorname{erf}\left(\frac{t}{\tau_c}\right)$.
It can be inferred that the limit of $\gamma(t) \rightarrow 0$ leads to the standard Ramsey protocol \cite{Sensing_RMP_2017}. However, in the presence of the environment, one has to choose the suitable $t$, which is indicated as $T_{dph}$, in order to achieve quantum sensing advantage even in the presence of the environment. These results are obtained for the nearest-neighbor setting \cite{matsuzaki2011}. Note that, even in the presence of the environment, careful selection of $T_{dph}$  significantly enhances the accuracy of the parameter estimation with variable-range interaction. Regarding the nature of the Hamiltonian, although quasi-long-range or long-range is immaterial to the quantum advantage in the presence of noise, it heavily depends on the non-Markovian nature of the noise. However, we surmise that the reason for the benefit in quasi-long-range interactions is again related to the generation of the spin-polarization wave.
\bibliography{metro_lr}

\end{document}